%% file: short_condmat_Tmatrix.tex
\documentclass[prb,twocolumn,showpacs,amsmath,amssymb,floatfix]{revtex4}
\usepackage{graphicx}
\usepackage{graphics}
\usepackage{dcolumn}
\usepackage{bm}
\usepackage{amsmath}
\usepackage{latexsym}
\usepackage{amsfonts}
\usepackage{amssymb}
\usepackage[rightcaption]{sidecap}
\usepackage{color}
\usepackage{float}
\makeatletter

\newcommand{\Rmnum}[1]{\expandafter\@slowromancap\romannumeral  #1@}

\makeatother
\begin{document}

\title{ Kadanoff-Baym equations  and approximate double occupancy in a Hubbard dimer.}
\date{\today}
\author{M. Puig von Friesen}
\author{C. Verdozzi}
\author{C.-O. Almbladh}
\affiliation{Mathematical Physics and European Theoretical Spectroscopy Facility (ETSF), Lund University, 22100  Lund, Sweden}
\begin{abstract}%
We use the Kadanoff-Baym equations within the framework of many-body perturbation theory to study
the double occupancies  $\langle \hat{n}_{R \uparrow} \hat{n}_{R \downarrow} \rangle$
in a Hubbard dimer. The double occupancies are obtained from the equations of motion of the 
single-particle Green's function. Our calculations show that the approximate double occupancies 
can become negative. This is a shortcoming of approximate, and yet conserving, many-body self-energies. 
Among the tested perturbation schemes, the $T$-matrix approximation is the only one providing double occupancies which are always positive.

\end{abstract}
\maketitle

This brief note reports of preliminary results on the use of the 
Kadanoff-Baym Equations (KBE) \cite{KadBay,Keldysh} to compute double 
occupancies. The case discussed here is that of a 
Hubbard dimer, described by the time-dependent Hamiltonian:
\begin{eqnarray}
\!\!H=\!\!\sum_\sigma [ -a^\dagger_{1\sigma} a_{2\sigma}-a^\dagger_{2\sigma} a_{1\sigma}  + w(t)\hat{n}_{1\sigma} ]\!+U\!\!\!\sum_{R=1,2} \!\!\hat{n}_{R\uparrow}\hat{n}_{R\downarrow},\;\;
\end{eqnarray}
where $U$ is the on-site and interaction,
$\hat{n}_{R\sigma}=a_{R\sigma}^{\dagger}a_{R\sigma}$ and $\sigma=\uparrow,\downarrow$. 
The term $w(t)$ is a time-dependent, spin-independent field; in $H$, all parameters are in units of the magnitude of the hopping term
(the latter is taken =-1). In the, $t=0$, initial state the dimer contains two electrons with opposite spin; this 
remains true at all times, since $H$ has no spin-flip terms.  

The study of this simple system reveals a shortcoming of the KBE when 
they are used together with many-body approximations (MBA:s). Following a route
based on the equations of motion of the single-particle Green's function, some 
conserving MBA:s (namely, the second Born approximation, BA, 
and the $GW$ approximation, GWA) may give negative double occupancies
 $\langle \hat{n}_{R \uparrow} \hat{n}_{R \downarrow} \rangle$,
which is clearly unphysical.
The results obtained with the $T$-matrix approximation (TMA) are exempt from this 
problem. This is no accident, and in fact we provide  an analytical proof of why, 
in the TMA, local double occupancies are intrinsically positive.\newline
{\it \bf Kadanoff-Baym equations. -} The KBE govern the time evolution of the 
non-equilibrium, two-time, single-particle Green's function 
$G(1,2)=-i\langle T_\gamma\left[\psi(1)\psi^\dagger(2)\right]\rangle$, where 
1 denote single particle space/spin and time labels, $r_1 \sigma_1 t_1$.
Here, $T_\gamma$ orders the times $t_1,t_2$ on the Keldysh contour\cite{Keldysh} $ \gamma$ and the field operators are in the Heisenberg picture; 
the brackets $\langle\rangle$ denote averaging over the initial state (or thermal equilibrium).
Showing explicitly only the time labels (matrix notation/multiplication in space and spin indices 
is adopted), and specializing to time $t_{1}$, we have  
$\left(i\partial_{t_{1}}-h\left(t_{1}\right)\right)G\left(t_{1},t_{2}\right)=
\delta(12)+\int_{\gamma}\Sigma\left(t_{1},t\right)G\left(t,t_{2}\right)dt$. 
Here $h$ is the single particle Hamiltonian and $\Sigma$, 
the kernel of the integral equation, is the self energy, which accounts for 
the interactions and is described within a given MBA, as discussed below. 
The initial state is the correlated ground state 
(we work at temperature, $1/\beta\rightarrow0$), obtained by solving the 
Dyson equation $G=G_{0}+G_{0}\Sigma[G] G$ self consistently, with $(\epsilon-h)G_0=1$.
In practice we solve these equations using an algorithm discussed in \cite{pva1}. 

The MBA:s we consider are all conserving \cite{KadBay}, i.e. quantities such as the total energy, 
the number of particles and momentum are constant during the time evolution. 
The self energy in the BA includes all diagrams up to second order. 
The GWA \cite{Hedin} amounts to add up all the bubble diagrams which give rise to the screened 
interaction, $W=U+UPW$, where $P(12)=G(12)G(21)$. In this case the self energy is $\Sigma(12)=G(12)W(12)$.
In the TMA \cite{TMA} one constructs the $T$ by adding up all the ladder diagrams, $T=\Phi-\Phi U T$, where $\Phi(12)=G(12)G(12)$. 
The expression of the self energy then becomes $\Sigma(12)=U^2G(21)T(12)$. For a detailed account
of the different MBA:s see e.g. \cite{pva1}. \newline
%
{\it \bf Double occupancy from the KBE. -} 
While in principle $G$ gives access only to expectation values
of one-body operators, its equations of motion (i.e., the KBE) permit to obtain also some quantities
which involve expectation values of two-body operators, such as, e.g., the total energy.
This route can be also used for the double occupancy $d_R$ \cite{Abrikosov};
for on-site $\left(U_R\neq0 \right)$ interactions, 
\begin{equation}
d_R=\left<\hat{n}_{R\uparrow}\hat{n}_{R\downarrow}\right>=-\frac{i}{U_R}\left[\int_\gamma \Sigma\left(13\right)G\left(31^{+}\right)d3\right]_{RR}.
\label{doccformula}\end{equation}
%
\begin{figure}[tbh]
\begin{center}
\includegraphics[width=7.7cm, clip=true]{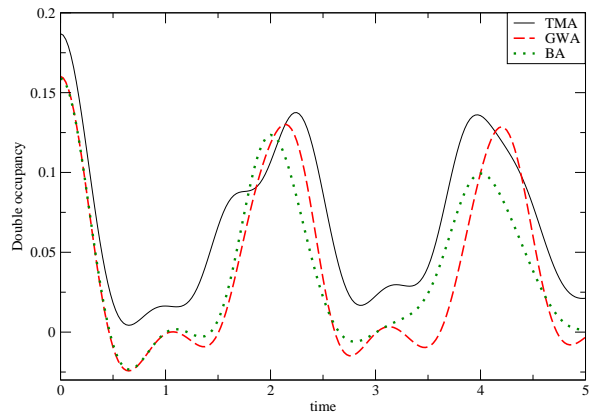}\\
\end{center}
\caption{Double occupancy.}
\label{doubleoccupancy}
\end{figure}
\noindent Fig. \ref{doubleoccupancy} shows the time-dependent double occupancy on site 1 
for $U=2$ and $w\left(t\right)=w_0\theta\left(t\right)$, where $w_0=3$. 
We see that the ground state ($t=0$) has positive double occupancies for 
all approximations while the BA and the GWA become negative after some time.
This unphysical behavior is a drawback that these two MBA:s can have.  
According to Fig.\ref{doubleoccupancy} the TMA, among the examined MBA:s, 
is the only one giving always positive correlation functions.
\input posCV_short_condmat.tex

We acknowledge discussions with Ulf von Barth. This work was supported by the 
EU 6th framework Network of Excellence NANOQUANTA (NMP4-CT-2004-500198) and the 
European Theoretical Spectroscopy Facility (INFRA-2007-211956).

\end{document}

%% file: posCV_short_condmat.tex
\newcommand{\vvr}{{\mathbf{r}}}

\noindent{\it \bf Positiveness of the \textit{T}-matrix. -} 
All the MBA:s used in this work are conserving. This, however,
does not guarantee that other properties (e.g. spectral 
features \cite{COAspectra, Holm} or response functions) obtained with
such MBA:s automatically
satisfy further basic criteria. In fact, for our dimer, 
density correlations and pair correlations for GWA and BA 
may violate the positiveness condition: 
$ \langle \psi^\dagger(1) \psi^\dagger(2) \psi(2) \psi(1) \rangle \ge 0$ .
The pair correlation function
is a rather sensitive measure of the quality of a MBA.
It has been known for a long time that the 
Random-Phase-Approximation (RPA) \cite{Lindhard}
in the ground state
gives pair correlations which, at metallic densities,  are strongly negative 
at short distances \cite{Alf}.  Later on, negative pair correlations at short 
distances were also found  within the self-consistent GWA \cite{Holm}.

Would the situation improve if the response function was
calculated self-consistently from \textit{changes} in $\Sigma$ and 
thereby via the Bethe-Salpeter equation?  Such response function fulfills 
all macroscopic conservation laws,  but not necessarily yields pair 
correlation functions everywhere positive. For example, the RPA discussed above
is the response function in the Kadanoff-Baym sense of the conserving
Hartree approximation but it violates the positiveness condition.

We now turn specifically to the TMA, to show why for this approximation
the pair correlation is manifestly positive. 
In the ground state, for the TMA, the pair correlation function
is a simple contraction of the $T$-matrix itself,
$\langle \hat{n}_{R\uparrow} \hat{n}_{R\downarrow}\rangle = -i T_{RR}(t, t^+)
= -i T^<_{RR}(t, t)$, (here $>$ ($<$) refers to the electron (hole) part)
and its  positiveness is a consequence of the positiveness
of the $T$-matrix spectral function.  The basic 
building block in the TMA is $\Phi_{RR^\prime} = -i G_{RR^\prime}(t-t^\prime) 
G_{RR^\prime}(t-t^\prime)$. In Fourier space we have
\begin{eqnarray}
\Phi_{RR^\prime}(\epsilon) = \int \frac{C_{RR^\prime}(\epsilon^\prime) d\epsilon^\prime}
{\epsilon - \epsilon^\prime + i \eta\; sgn(\epsilon^\prime - 2 \mu)} ,
\label{eq:C_spec}
\end{eqnarray}
where the spectral function $C_{RR^\prime}$ is given by 
\begin{eqnarray}
C_{RR^\prime}(\epsilon) =  \int_{\epsilon - \mu}^\mu 
A_{RR^\prime}(\epsilon^\prime) A_{RR^\prime}(\epsilon - \epsilon^\prime) d\epsilon^\prime.
\nonumber
\end{eqnarray}
\noindent Thus, $C_{RR^\prime}(\epsilon)$ is positive (negative) definite for $\epsilon < 2 \mu$ 
($\epsilon > 2 \mu$).
The Dyson equation for the $T$-matrix is
$ \hat{T}(\epsilon) = \hat{\Phi}(\epsilon) - \hat{\Phi}(\epsilon)\hat{U}
\hat{T}(\epsilon) $
(here $\hat{\Phi}, \hat{T}, \hat{U}$ are matrices in site indices, 
and $U_{RR'}=U_R\delta_{RR'}$). This gives
\begin{eqnarray*}
&&\hat{T}^\lessgtr = \hat{\Phi}^\lessgtr - \hat{\Phi}^\lessgtr \hat{U} \hat{T}^a -
\hat{\Phi}^r \hat{U} \hat{T}^\lessgtr , \\
&&(1 + \hat{\Phi}^r \hat{U}) \hat{T}^\lessgtr = \hat{\Phi}^\lessgtr (
1 - \hat{U} \hat{T}^a ),
\end{eqnarray*}
and
\begin{eqnarray}
\hat{T}(\epsilon) - \hat{T}^\dagger(\epsilon)
= [1 - \hat{T^r}(\epsilon) \hat{U} ] 
[ \hat{\Phi}(\epsilon) - \hat{\Phi}^\dagger(\epsilon) ]
[1 - \hat{U} \hat{T}^a(\epsilon) ],
\nonumber
\end{eqnarray}
where we have used the identity
$(1 - \hat{U} \hat{T} )  (1 + \hat{\Phi} \hat{U}) = 1$.

\noindent The $T$-matrix has thus the spectral decomposition
\begin{eqnarray}
\hat{T}(\epsilon) = \int \frac{\hat{D}(\epsilon^\prime) d\epsilon^\prime}
{\epsilon - \epsilon^\prime + i \eta\; sgn(\epsilon^\prime - 2 \mu)}
\label{eq:T_D}, 
\end{eqnarray}
where  $\hat{D}(\epsilon) = [1 - \hat{T}^r(\epsilon) \hat{U} ] \hat{C}(\epsilon)
[1 - \hat{U} \hat{T}^a(\epsilon) ]$. 
Consequently, $\hat{D}(\epsilon)$ is positive definite for $\epsilon < 2 \mu$,
and
\begin{eqnarray}
\langle \hat{n}_{R\uparrow} \hat{n}_{R\downarrow}\rangle = -i T_{RR}(t, t^+)
= \int_{-\infty}^{2 \mu} D_{RR}(\epsilon^\prime) d\epsilon^\prime  \ge 0 .
\label{eq:pos_T_hubb}
\end{eqnarray}
\noindent The above result remains valid i) for a general two-body interaction
$u(\vvr_1 - \vvr_2) \delta(t_1 - t_2)$, provided we use 
a symmetrized TMA which includes both direct and exchange ladder
diagrams; ii)   in the presence of an external field for $t > 0$.
In this case, the proof is similar to the proof that manifestly positive 
$\mp i \Sigma^\lessgtr$ give manifestly positive $\mp i G^\lessgtr$ via
the KBE \cite{inpreparation}. 